\documentclass[11pt]{article}
\usepackage{amssymb,amsmath,amsfonts}
\usepackage{graphicx}
\usepackage{graphics}
\usepackage{eepic,epsfig}

\begin{document}

\title{Electromagnetic Casimir densities induced by a conducting cylindrical
shell in the cosmic string spacetime }
\author{E. R. Bezerra de Mello$^{1}$\thanks{%
E-mail: emello@fisica.ufpb.br},\thinspace\ V. B. Bezerra$^{1}$\thanks{%
E-mail: valdir@fisica.ufpb.br}, \thinspace\ A. A. Saharian$^{1,2}$\thanks{%
E-mail: saharyan@server.physdep.r.am} \\
\\
\textit{$^{1}$Departamento de F\'{\i}sica, Universidade Federal da Para\'{\i}%
ba}\\
\textit{58.059-970, Caixa Postal 5.008, Jo\~{a}o Pessoa, PB, Brazil}\vspace{%
0.3cm}\\
\textit{$^2$Department of Physics, Yerevan State University,}\\
\textit{375025 Yerevan, Armenia}}
\maketitle

\begin{abstract}
We investigate the renormalized vacuum expectation values of the field
square and the energy-momentum tensor for the electromagnetic field inside
and outside of a conducting cylindrical shell in the cosmic string
spacetime. By using the generalized Abel-Plana formula, the vacuum
expectation values are presented in the form of the sum of boundary-free and
boundary-induced parts. The asymptotic behavior of the vacuum expectation
values of the field square, energy density and stresses are investigated in
various limiting cases.
\end{abstract}

\bigskip


\bigskip

\section{Introduction}

Symmetry braking phase transitions in the early universe have several
cosmological consequences and provide an important link between particle
physics and cosmology. In particular, within the framework of grand unified
theories various types of topological defects are predicted to be formed
\cite{Vile85}. Among them the cosmic strings are of special interest.
Although the recent observational data on the cosmic microwave background
radiation have ruled out cosmic strings as the primary source for primordial
density perturbations, they are still candidates for the generation of a
number of interesting physical effects such as the generation of
gravitational waves and gamma ray bursts. Recently, cosmic strings attract a
renewed interest partly because a variant of their formation mechanism is
proposed in the framework of brane inflation \cite{Sara02}.

In the simplest theoretical model describing the infinite straight cosmic
string the spacetime is locally flat except on the string where it has a
delta shaped curvature tensor. In quantum field theory the corresponding
non-trivial topology induces non-zero vacuum expectation values for physical
observables. Explicit calculations for the geometry of a single cosmic
string have been done for different fields \cite{Hell86}-\cite{Beze06}.
Another type of vacuum polarization arises when boundaries are present. The
imposed boundary conditions on quantum fields alter the zero-point
fluctuations spectrum and result in additional shifts in the vacuum
expectation values of physical quantities. This is the well-known Casimir
effect (for a review see \cite{Most97}). In a previous paper \cite{Beze06sc}%
, we have studied the configuration with both types of sources for the
polarization of the scalar vacuum, namely, a cylindrical boundary and a
cosmic string, assuming that the boundary is coaxial with the string and
that on this surface the scalar field obeys Robin boundary condition. For a
massive scalar field with an arbitrary curvature coupling parameter we
evaluated the Wightman function and the vacuum expectation values of the
field square and the energy-momentum tensor. (For a combination of
topological and boundary-induced quantum effects in the gravitational field
of the global monopole see Refs. \cite{Saha03Mon}.)

In the present paper we consider the polarization of the electromagnetic
vacuum by a perfectly conducting cylindrical shell coaxial with the cosmic
string. Specifically, we evaluate the vacuum expectation values of the field
square and the energy-momentum tensor in both interior and exterior regions
of the shell. In addition to describing the physical structure of the
quantum field at a given point, the energy-momentum tensor acts as a source
of gravity in the Einstein equations. It therefore plays an important role
in modelling a self-consistent dynamics involving the gravitational field.
In the problem under consideration all calculations can be performed in a
closed form and it constitutes an example in which the gravitational and
boundary-induced polarizations of the vacuum can be separated in different
contributions. It is worth calling attention to the fact that the
renormalized vacuum expectation value of the energy-momentum tensor for the
electromagnetic field in the geometry of a cosmic string without boundaries
is evaluated in \cite{Frol87,Dowk87b} and the effects induced by the core
internal structure are investigated in \cite{Alle92}.

In \cite{Witt85} it has been shown that under certain conditions strings
behave like superconducting wires whose passage through galactic magnetic
fields may generate a number of interesting effects (see also \cite{Vile85}%
). In particular, this type of strings might be observable as sources for
synchrotron radiation and cosmic rays. These effects are of special interest
for light strings which would not be observed by their gravitational
interactions. From the point of view of the physics in the region outside
the string core, the geometry considered in the present paper can be viewed
as a simplified model for the superconducting string, in which the string
core in what concerns its superconducting effects is taken to be an ideal
conductor. This model presents a framework in which the influence of the
finite core effects on physical processes in the vicinity of the cosmic
string can be investigated. In particular, it enables to specify conditions
under which the idealized model with the core of zero thickness can be used.
The corresponding results may shed light \ upon features of finite core
effects in more realistic models, including those used for string-like
defects in crystals and superfluid helium. In addition, the problem
considered here is of interest as an example with combined topological and
boundary induced quantum effects in which the vacuum characteristics such as
energy density and stresses can be found in closed analytic form.

We have organized the paper as follows. In the next section we consider the
vacuum expectation values of the field square and the energy-momentum tensor
in the region inside the cylindrical shell. By using the generalized
Abel-Plana formula, we extract from the mode-sums the parts corresponding to
the geometry of a cosmic string without the shell. The parts induced by the
shell are investigated in various asymptotic regions for the parameters. The
vacuum expectation values in the exterior region are studied in section \ref%
{sec:exter}. Finally, in section \ref{sec:Conc} the main results are
summarized.

\section{Vacuum expectation values inside a cylindrical shell}

\label{sec:Inter}

We consider the background spacetime corresponding to an infinitely long
straight cosmic string with the conical line element%
\begin{equation}
ds^{2}=dt^{2}-dr^{2}-r^{2}d\phi ^{2}-dz{}^{2},  \label{ds21}
\end{equation}%
where $0\leqslant \phi \leqslant \phi _{0}$ and the spatial points $(r,\phi
,z)$ and $(r,\phi +\phi _{0},z)$ are to be identified. The planar angle
deficit is related to the mass per unit length of the string, $\mu _{0}$, by
$2\pi -\phi _{0}=8\pi G\mu _{0}$, where $G$ is the Newton gravitational
constant. In this paper we are interested in the change of the vacuum
expectation values (VEVs) of the field square and the energy-momentum tensor
for the electromagnetic field, induced by a conducting cylindrical shell
coaxial \ with the string. Expanding the field operator in terms of the
creation and annihilation operators, it can be seen that the VEV for a
quantity $F\left\{ A_{i},A_{k}\right\} $ bilinear in the field can be
written in the form of the mode-sum%
\begin{equation}
\langle 0|F\left\{ A_{i},A_{k}\right\} |0\rangle =\sum_{\alpha }F\left\{
A_{\alpha i},A_{\alpha k}^{\ast }\right\} ,  \label{VEVbil}
\end{equation}%
where $\left\{ A_{\alpha i},A_{\alpha k}^{\ast }\right\} $ is a complete set
of solutions of the classical field equations specified by a set of quantum
numbers $\alpha $.

In accordance with (\ref{VEVbil}), to evaluate the VEVs for the square of
the electric and magnetic fields and the energy-momentum tensor, we need the
corresponding eigenfunctions. Firstly, let us consider the region inside the
cylindrical shell. For the geometry under consideration we have two
different eigenfunctions corresponding to the waves of the electric and
magnetic types. In the Coulomb gauge, the vector potentials for these waves
are given by the formulae%
\begin{equation}
\mathbf{A}_{\alpha }=\beta _{\alpha }\left\{
\begin{array}{cc}
(1/i\omega )\left( \gamma ^{2}\mathbf{e}_{3}+ik\nabla _{t}\right)
J_{q|m|}(\gamma r)\exp \left[ i\left( qm\phi +kz-\omega t\right) \right] , &
\lambda =0 \\
-\mathbf{e}_{3}\times \nabla _{t}\left\{ J_{q|m|}(\gamma r)\exp \left[
i\left( qm\phi +kz-\omega t\right) \right] \right\} , & \lambda =1%
\end{array}%
\right. ,  \label{Aalpha}
\end{equation}%
where $\mathbf{e}_{3}$ is the unit vector along the cosmic string, $\nabla
_{t}$ is the part of the nabla operator transverse to the string, $J_{\nu
}(x)$ is the Bessel function, and%
\begin{equation}
\omega ^{2}=\gamma ^{2}+k^{2},\;q=2\pi /\phi _{0},\;m=0,\pm 1,\pm 2,\ldots .
\label{omega}
\end{equation}%
Here and in what follows $\lambda =0$ and $\lambda =1$ correspond to the
cylindrical waves of the electric (transverse magnetic (TM)) and magnetic
(transverse electric (TE)) types, respectively. The normalization
coefficient in (\ref{Aalpha}) is found from the orthonormalization condition
for the vector potential:%
\begin{equation}
\int dV\,\mathbf{A}_{\alpha }\cdot \mathbf{A}_{\alpha ^{\prime }}^{\ast }=%
\frac{2\pi }{\omega }\delta _{\alpha \alpha ^{\prime }},  \label{Anorm}
\end{equation}%
where the integration goes over the region inside the shell. From this
condition, by using the standard integral involving the square of the Bessel
function, one finds%
\begin{equation}
\beta _{\alpha }^{2}=\frac{qT_{q|m|}(\gamma a)}{\pi \omega a\gamma },
\label{betalf}
\end{equation}%
where we have introduced the notation
\begin{equation}
T_{\nu }(x)=x\left[ J_{\nu }^{^{\prime }2}(x)+(1-\nu ^{2}/x^{2})J_{\nu
}^{2}(x)\right] ^{-1}.  \label{Tnux}
\end{equation}

The eigenvalues for the quantum number $\gamma $ are determined by the
standard boundary conditions for the electric and magnetic fields on the
cylindrical shell, namely $\mathbf{n}\times \mathbf{E}=0$ and $\mathbf{n}%
\cdot \mathbf{B}=0$, with $\mathbf{n}$\ being the normal to the boundary.
From these boundary conditions we see that these eigenvalues are solutions
of the equation
\begin{equation}
J_{q|m|}^{(\lambda )}(\gamma a)=0, \quad \lambda =0,1,  \label{modes1}
\end{equation}%
where $a$ is the radius of the cylindrical shell, $J_{\nu }^{(0)}(x)=J_{\nu
}(x)$ and $J_{\nu }^{(1)}(x)=J^{\prime}_{\nu }(x)$. We will denote the
corresponding eigenmodes by $\gamma a=j_{m,n}^{(\lambda )}$, $n=1,2,\ldots $%
. As a result the eigenfunctions are specified by the set of quantum numbers
$\alpha =(k,m,\lambda ,n)$.

\subsection{Vacuum expectation values of the field square}

\label{subsec:F2}

Substituting the eigenfunctions into the corresponding mode-sum formula, for
the VEVs of the squares of the electric and magnetic fields inside the
shell, we find%
\begin{equation}
\langle 0|F^{2}|0\rangle =\sum_{\alpha }\mathbf{F}_{\alpha }\cdot \mathbf{F}%
_{\alpha }^{\ast }=\frac{2q}{\pi a^{3}}\sideset{}{'}{\sum}_{m=0}^{\infty
}\int_{-\infty }^{+\infty }dk\sum_{\lambda =0,1}\sum_{n=1}^{\infty }\frac{%
j_{m,n}^{(\lambda )3}T_{qm}(j_{m,n}^{(\lambda )})}{\sqrt{j_{m,n}^{(\lambda
)2}+k^{2}a^{2}}}g_{qm}^{(\eta _{F\lambda })}\left[ k,J_{qm}(j_{m,n}^{(%
\lambda )}r/a)\right] ,  \label{F2}
\end{equation}%
where $F=E,B$ with $\eta _{E\lambda }=\lambda $, $\eta _{B\lambda
}=1-\lambda $, and the prime in the summation means that the term $m=0$
should be halved. In (\ref{F2}), for a given function $f(x)$, we have used
the notation
\begin{equation}
g_{\nu }^{(j)}\left[ k,f(x)\right] =\left\{
\begin{array}{cc}
(k^{2}r^{2}/x^{2})\left[ f^{\prime 2}(x)+\nu ^{2}f^{2}(x)/x^{2}\right]
+f^{2}(x), & j=0 \\
(1+k^{2}r^{2}/x^{2})\left[ f^{\prime 2}(x)+\nu ^{2}f^{2}(x)/x^{2}\right] , &
j=1%
\end{array}%
\right. .  \label{gnulam}
\end{equation}%
The expressions (\ref{F2}) corresponding to the electric and magnetic fields
are divergent. They may be regularized introducing a cutoff function $\psi
_{\mu }(\omega )$ with the cutting parameter $\mu $ which makes the
divergent expressions finite and satisfies the condition $\psi _{\mu
}(\omega )\rightarrow 1$ for $\mu \rightarrow 0$. After the renormalization
the cutoff function is removed by taking the limit $\mu \rightarrow 0$. An
alternative way is to consider the product of the fields at different
spacetime points and take the coincidence limit after the subtraction of the
corresponding Minkowskian part. Here we will follow the first approach.

In order to evaluate the mode-sum in (\ref{F2}), we apply to the series over
$n$ the generalized Abel-Plana summation formula \cite{Saha87}
\begin{eqnarray}
\sum_{n=1}^{\infty }T_{qm}(j_{m,n}^{(\lambda )})f(j_{m,n}^{(\lambda )}) &=&%
\frac{1}{2}\int_{0}^{\infty }dx\,f(x)+\frac{\pi }{4}\underset{z=0}{\mathrm{%
Res}} f(z)\frac{Y_{qm}^{(\lambda )}(z)}{J_{qm}^{(\lambda )}(z)}  \notag \\
&&-\frac{1}{2\pi }\int_{0}^{\infty }dx\,\frac{K_{qm}^{(\lambda )}(x)}{%
I_{qm}^{(\lambda )}(x)}\left[ e^{-qm\pi i}f(ix)+e^{qm\pi i}f(-ix)\right] ,
\label{sumformula}
\end{eqnarray}%
where $Y_{\nu }(z)$ is the Neumann function and $I_{\nu }(z)$, $K_{\nu }(z)$
are the modified Bessel functions. As it can be seen, for points away from
the shell the contribution to the VEVs coming from the second integral term
on the right-hand side of (\ref{sumformula}) is finite in the limit $\mu
\rightarrow 0$ and, hence, the cutoff function in this term can be safely
removed. As a result the VEVs can be written as%
\begin{equation}
\langle 0|F^{2}|0\rangle =\langle 0_{s}|F^{2}|0_{s}\rangle +\left\langle
F^{2}\right\rangle _{\mathrm{b}},  \label{F21}
\end{equation}%
where%
\begin{eqnarray}
\langle 0_{s}|F^{2}|0_{s}\rangle &=&\frac{q}{\pi }\sideset{}{'}{\sum}%
_{m=0}^{\infty }\int_{-\infty }^{+\infty }dk\int_{0}^{\infty }d\gamma \,%
\frac{\gamma ^{3}\psi _{\mu }(\omega )}{\sqrt{\gamma ^{2}+k^{2}}}  \notag \\
&&\left[ \left( 1+2\frac{k^{2}}{\gamma ^{2}}\right) \left( J_{qm}^{\prime
2}(\gamma r)+\frac{q^{2}m^{2}}{\gamma ^{2}r^{2}}J_{qm}^{2}(\gamma r)\right)
+J_{qm}^{2}(\gamma r)\right] ,  \label{F2s}
\end{eqnarray}%
and%
\begin{equation}
\langle F^{2}\rangle _{\mathrm{b}}=\frac{4q}{\pi ^{2}}\sideset{}{'}{\sum}%
_{m=0}^{\infty }\int_{0}^{\infty }dk\sum_{\lambda =0,1}\int_{k}^{\infty
}dx\,x^{3}\,\frac{K_{qm}^{(\lambda )}(xa)}{I_{qm}^{(\lambda )}(xa)}\frac{%
G_{qm}^{(\eta _{F\lambda })}\left[ k,I_{qm}(xr)\right] }{\sqrt{x^{2}-k^{2}}}.
\label{F2b0}
\end{equation}%
Note that in Eq. (\ref{F2b0}) we used the notations%
\begin{equation}
G_{\nu }^{(j)}\left[ k,f(x)\right] =\left\{
\begin{array}{cc}
(k^{2}r^{2}/x^{2})\left[ f^{\prime 2}(x)+\nu ^{2}f^{2}(x)/x^{2}\right]
+f^{2}(x), & j=0 \\
(k^{2}r^{2}/x^{2}-1)\left[ f^{\prime 2}(x)+\nu ^{2}f^{2}(x)/x^{2}\right] , &
j=1%
\end{array}%
\right. .  \label{Gnuj}
\end{equation}%
The second term on the right-hand side of Eq. (\ref{F21}) vanishes in the
limit $a\rightarrow \infty $. It is worth calling attention to the fact that
this term depends not only on the boundary but also on the presence of the
cosmic string. On the other hand the first one does not depend on $a$. Thus,
we can conclude that the term $\langle 0_{s}|F^{2}|0_{s}\rangle $
corresponds to the part in VEVs\ when the cylindrical shell is absent with
the corresponding vacuum state $|0_{s}\rangle $. Note that for the geometry
without boundaries one has $\langle 0_{s}|E^{2}|0_{s}\rangle =\langle
0_{s}|B^{2}|0_{s}\rangle $. Hence, the application of the generalized
Abel-Plana formula enables us to extract from the VEVs the boundary-free
parts and to write the boundary-induced parts in terms of the exponentially
convergent integrals.

Now, let us calculate the boundary-free part contribution. To do this, let
us firstly consider the case when the parameter $q$ is an integer. In this
case the summation over $m$ can be done by using the formula \cite%
{Davi88,Prud86}
\begin{equation}
\sideset{}{'}{\sum}_{m=0}^{\infty }J_{qm}^{2}(y)=\frac{1}{2q}%
\sum_{l=0}^{q-1}J_{0}(2y\sin (\pi l/q)).  \label{rel1}
\end{equation}%
Differentiating this relation and using the equation for the Bessel
function, it can be seen that%
\begin{equation}
\sideset{}{'}{\sum}_{m=0}^{\infty }\left[ J_{qm}^{^{\prime }2}(y)+\frac{%
q^{2}m^{2}}{y^{2}}J_{qm}^{2}(y)\right] =\sum_{l=0}^{q-1}\frac{\cos (2\pi l/q)%
}{2q}J_{0}(2y\sin (\pi l/q)).  \label{rel2}
\end{equation}%
Thus, using this result, we find that the corresponding VEV is given by%
\begin{eqnarray}
\langle 0_{s}|F^{2}|0_{s}\rangle &=&\frac{1}{2\pi }\sum_{l=0}^{q-1}\int_{-%
\infty }^{+\infty }dk\int_{0}^{\infty }d\gamma \,\frac{\gamma \psi _{\mu
}(\omega )}{\sqrt{\gamma ^{2}+k^{2}}}  \notag \\
&&\times J_{0}(2\gamma r\sin (\pi l/q))\left[ \left( \gamma
^{2}+2k^{2}\right) \cos (2\pi l/q)+\gamma ^{2}\right] .  \label{F2s1}
\end{eqnarray}%
The $l=0$ term in this formula corresponds to the VEV in the Minkowski bulk,
$\langle 0_{M}|F^{2}|0_{M}\rangle $. Therefore, the renormalized value of
the VEV is obtained subtracting this term, that is%
\begin{equation}
\langle F^{2}\rangle _{s,\mathrm{ren}}=\lim_{\mu \rightarrow 0}\left[
\langle 0_{s}|F^{2}|0_{s}\rangle -\langle 0_{M}|F^{2}|0_{M}\rangle \right] .
\label{F2sren}
\end{equation}%
Introducing polar coordinates $(\omega ,\theta )$ in the $(\gamma ,k)$
plane, the $\omega $-integral with the cutoff function $\psi _{\mu }(\omega
)=e^{-\mu \omega }$ is evaluated by using the formula \cite{Prud86}%
\begin{equation}
\int_{0}^{\infty }d\omega \,\omega ^{n}e^{-\mu \omega }J_{0}(\omega
x)=(-1)^{n}\frac{\partial ^{n}}{\partial \mu ^{n}}\frac{1}{\sqrt{\mu
^{2}+x^{2}}},  \label{intform4}
\end{equation}%
which leads to the result%
\begin{equation}
\langle F^{2}\rangle _{s,\mathrm{ren}}=-\frac{1}{\pi r}\sum_{l=1}^{q-1}\frac{%
1}{y_{l}}\lim_{\mu \rightarrow 0}\left( \frac{\partial ^{3}}{\partial \mu
^{3}}\int_{0}^{1}dx\,\frac{y_{l}^{2}-1-x^{2}}{\sqrt{\mu
^{2}y_{l}^{2}/4r^{2}+1-x^{2}}}\right) ,  \label{F2sren2n}
\end{equation}%
where $y_{l}=1/\sin (\pi l/q)$. Thus, evaluating the integral and taking the
limit we find%
\begin{equation}
\langle F^{2}\rangle _{s,\mathrm{ren}}=-\frac{1}{4\pi r^{4}}%
\sum_{l=1}^{q-1}y_{l}^{4}=-\frac{(q^{2}-1)(q^{2}+11)}{180\pi r^{4}}.
\label{F2sren3}
\end{equation}%
This result is an analytic function of $q$ and, hence, by analytic
continuation it is also valid for non-integer values of $q$.

Now, we turn to the investigation of the boundary-induced part given by (\ref%
{F2b0}). Changing the integration variable to $y=\sqrt{x^{2}-k^{2}}$ and
introducing polar coordinates in the $(k,y)$ plane, after the explicit
integration over the angular part, one obtains%
\begin{equation}
\int_{0}^{\infty }dk\,k^{m}\int_{k}^{\infty }dx\,\frac{xf(x)}{\sqrt{%
x^{2}-k^{2}}}=\frac{\sqrt{\pi }\Gamma \left( \frac{m+1}{2}\right) }{2\Gamma
\left( \frac{m}{2}+1\right) }\int_{0}^{\infty }dx\,x^{m+1}f(x).
\label{Hash1}
\end{equation}%
By using this formula, the part in the VEV induced by the cylindrical shell
can be written in the form
\begin{equation}
\langle F^{2}\rangle _{\mathrm{b}}=\frac{q}{\pi }\sideset{}{'}{\sum}%
_{m=0}^{\infty }\sum_{\lambda =0,1}\int_{0}^{\infty }dx\,x^{3}\frac{%
K_{qm}^{(\lambda )}(xa)}{I_{qm}^{(\lambda )}(xa)}G_{qm}^{(\eta _{F\lambda })}%
\left[ I_{qm}(xr)\right] ,  \label{F2b}
\end{equation}%
where we have used the notation%
\begin{equation}
G_{\nu }^{(j)}\left[ f(x)\right] =\left\{
\begin{array}{cc}
f^{\prime 2}(x)+\nu ^{2}f^{2}(x)/x^{2}+2f^{2}(x), & j=0 \\
-f^{\prime 2}(x)-\nu ^{2}f^{2}(x)/x^{2}, & j=1%
\end{array}%
\right. .  \label{Gnujtilde}
\end{equation}%
This result tells us that the boundary-induced parts for the electric and
magnetic fields are different and, hence, the presence of the shell breaks
the electric-magnetic symmetry in the VEVs. Of course, this is a consequence
of different boundary conditions for the electric and magnetic fields.

The expression in the right-hand side of (\ref{F2b}) is finite for $0<r<a$
and diverges on the shell. To find the leading term in the corresponding
asymptotic expansion, we note that near the surface the main contribution
comes from large values of $m$. By using the uniform asymptotic expansions
of the modified Bessel functions (see, for instance, \cite{hand}) for large
values of the order, up to the leading order, we find%
\begin{equation}
\langle E^{2}\rangle _{\mathrm{b}}\approx -\langle B^{2}\rangle _{\mathrm{b}%
}\approx \frac{3}{4\pi (a-r)^{4}}.  \label{E2binnear}
\end{equation}%
This leading term does not depend on the angle deficit and has opposite
signs for the electric and magnetic fields. In particular, it is cancelled
in the evaluation of the vacuum energy density. Surface divergences
originate in the unphysical nature of perfect conductor boundary conditions
and are well-known in quantum field theory with boundaries. In reality the
expectation values will attain a limiting value on the conductor surface,
which will depend on the molecular details of the conductor. From the
formulae given above it follows that the main contribution to $\langle
F^{2}\rangle _{\mathrm{b}}$ are due to the frequencies $\omega \lesssim
(a-r)^{-1}$. Hence, we expect that formula (\ref{F2b}) is valid for real
conductors up to distances $r$ for which $(a-r)^{-1}\ll \omega _{0}$, with $%
\omega _{0}$ being the characteristic frequency, such that for $\omega
>\omega _{0}$ the conditions for perfect conductivity fail.

Near the string, $r/a\ll 1$, the asymptotic behavior of the boundary induced
part in the VEVs of the field squares depends on the parameter $q$. For $%
q\geqslant 1$, the dominant contribution comes from the lowest mode $m=0$
and to the leading order one has%
\begin{eqnarray}
\langle E^{2}\rangle _{\mathrm{b}} &\approx &\frac{q}{\pi a^{4}}%
\int_{0}^{\infty }dx\,x^{3}\frac{K_{0}(x)}{I_{0}(x)}\approx 0.320\frac{q}{%
a^{4}},\;  \label{E2bnearcent} \\
\langle B^{2}\rangle _{\mathrm{b}} &\approx &-\frac{q}{\pi a^{4}}%
\int_{0}^{\infty }dx\,x^{3}\frac{K_{1}(x)}{I_{1}(x)}\approx -0.742\frac{q}{%
a^{4}}.  \label{B2bnearcent}
\end{eqnarray}%
For $q<1$ the main contribution comes form the mode with $m=1$ and the
boundary induced parts diverge on the string. The leading terms are given by%
\begin{equation}
\langle E^{2}\rangle _{\mathrm{b}}\approx -\langle B^{2}\rangle _{\mathrm{b}%
}\approx \frac{(r/a)^{2(q-1)}}{2^{2(q-1)}\pi \Gamma ^{2}(q)a^{4}}%
\int_{0}^{\infty }dx\,x^{2q+1}\left[ \frac{K_{q}(x)}{I_{q}(x)}-\frac{%
K_{q}^{\prime }(x)}{I_{q}^{\prime }(x)}\right] .  \label{E2bnearcent2}
\end{equation}%
As for the points near the shell, here the leading divergence is cancelled
in the evaluation of the vacuum energy density. In accordance with (\ref%
{F2sren3}), near the string the total VEV is dominated by the boundary-free
part. Here we have considered the VEV for the field square. The VEVs for the
bilinear products of the fields at different spacetime points may be
evaluated in a similar way.

Now, we turn to the investigation of the behavior of the boundary-induced
VEVs in the asymptotic regions of the parameter $q$. For small values of
this parameter, $q\ll 1$, the main contribution into (\ref{F2b}) comes from
large values of $m$. In this case, we can replace the summation over $m$ by
an integration in accordance with the correspondence%
\begin{equation}
\sideset{}{'}{\sum}_{m=0}^{\infty }h(qm)\rightarrow \frac{1}{q}%
\int_{0}^{\infty }dx\,h(x).  \label{replSumInt}
\end{equation}%
By making this replacement, we can see from (\ref{F2b}) that, in this
situation, the boundary induced VEVs tend to a finite value. Note that the
same is the case for the boundary-free part (\ref{F2sren3}). In the limit $%
q\gg 1$, the order of the modified Bessel functions is large for $m\neq 0$.
By using the corresponding asymptotic formulae it can be seen that the
contribution of these term is suppressed by the factor $\exp [-2qm\ln (a/r)]$%
. As a result, the main contribution comes from the lowest mode $m=0$ and
the boundary induced VEVs behave like $q$.

\subsection{Vacuum expectation value for the energy-momentum tensor}

\label{subsec:EMTint}

In this subsection we consider the vacuum energy-momentum tensor in the
region inside the cylindrical shell. Substituting the eigenfunctions (\ref%
{Aalpha}) into the corresponding mode-sum formula, we obtain (no summation
over $i$)%
\begin{equation}
\langle 0|T_{i}^{k}|0\rangle =\frac{q\delta _{i}^{k}}{4\pi ^{2}a^{3}}%
\sideset{}{'}{\sum}_{m=0}^{\infty }\int_{-\infty }^{+\infty }dk\sum_{\lambda
=0,1}\sum_{n=1}^{\infty }\frac{j_{m,n}^{(\lambda )3}T_{qm}(j_{m,n}^{(\lambda
)})}{\sqrt{j_{m,n}^{(\lambda )2}+k^{2}a^{2}}}f_{qm}^{(i)}\left[
k,J_{qm}(j_{m,n}^{(\lambda )}r/a)\right] ,  \label{Tik}
\end{equation}%
where we have introduced the notations%
\begin{eqnarray}
f_{\nu }^{(i)}\left[ k,f(x)\right] &=&(-1)^{i}\left( 2k^{2}/\gamma
^{2}+1\right) \left[ f^{\prime 2}(x)+\nu ^{2}f^{2}(x)/y^{2}\right]
+f^{2}(x),\;i=0,3,  \label{f0} \\
f_{\nu }^{(i)}\left[ k,f(x)\right] &=&(-1)^{i}f^{\prime 2}(x)-\left[
1+(-1)^{i}\nu ^{2}/x^{2}\right] f^{2}(x),\;i=1,2.  \label{fi}
\end{eqnarray}%
As in the case of the field square, we apply the summation formula (\ref%
{sumformula}) to rewrite the sum over $m$. This enables us to present the
VEV in the form of the boundary-free and boundary-induced parts as follows
\begin{equation}
\langle 0|T_{i}^{k}|0\rangle =\langle 0_{s}|T_{i}^{k}|0_{s}\rangle +\langle
T_{i}^{k}\rangle _{\mathrm{b}}.  \label{Tik1}
\end{equation}%
By taking into account (\ref{Hash1}), the part induced by the cylindrical
shell may be written in the form (no summation over $i$)
\begin{equation}
\langle T_{i}^{k}\rangle _{\mathrm{b}}=\frac{q\delta _{i}^{k}}{4\pi ^{2}}%
\sideset{}{'}{\sum}_{m=0}^{\infty }\sum_{\lambda =0,1}\int_{0}^{\infty
}dxx^{3}\frac{K_{qm}^{(\lambda )}(xa)}{I_{qm}^{(\lambda )}(xa)}F_{qm}^{(i)}%
\left[ I_{qm}(xr)\right] ,  \label{Tikb}
\end{equation}%
with the notations
\begin{eqnarray}
F_{\nu }^{(0)}[f(y)] &=&F_{\nu }^{(3)}[f(y)]=f^{2}(y),  \label{Fnu0} \\
F_{\nu }^{(i)}[f(y)] &=&-(-1)^{i}f^{\prime 2}(y)-\left[ 1-(-1)^{i}\nu
^{2}/y^{2}\right] f^{2}(y),\;i=1,2.  \label{Fnui}
\end{eqnarray}%
As it can be easily checked, this tensor is traceless and satisfies the
covariant continuity equation $\nabla _{k}\langle T_{i}^{k}\rangle _{\mathrm{%
b}}=0$. By using the inequalities $[I_{\nu }(y)K_{\nu }(y)]^{\prime }<0$ and
$I_{\nu }^{\prime }(y)<\sqrt{1+\nu ^{2}/y^{2}}I_{\nu }(y)$, and the
recurrence relations for the modified Bessel functions, it can be seen that
the boundary-induced parts in the vacuum energy density and axial stress are
negative, whereas the corresponding radial and azimuthal stresses are
positive.

The renormalized VEV of the energy-momentum tensor for the geometry without
the cylindrical shell is obtained by using the corresponding formulae for
the field square (\ref{F2sren3}). For the corresponding energy density one
finds \cite{Frol87,Dowk87b}
\begin{equation}
\langle T_{0}^{0}\rangle _{s,\mathrm{ren}}=\frac{1}{8\pi }\left( \langle
E^{2}\rangle _{s,\mathrm{ren}}+\langle B^{2}\rangle _{s,\mathrm{ren}}\right)
=-\frac{(q^{2}-1)(q^{2}+11)}{720\pi ^{2}r^{4}}.  \label{T00s}
\end{equation}%
Other components are found from the tracelessness condition and the
continuity equation.

Now, let us discuss the behavior of the boundary-induced part in the VEV of
the energy-momentum tensor in the asymptotic region of the parameters. Near
the cylindrical shell the main contribution comes from large values of $m$.
Thus, using the uniform asymptotic expansions for the modified Bessel
functions for large values of the order, up to the leading order, we find%
\begin{equation}
\langle T_{0}^{0}\rangle _{\mathrm{b}}\approx -\frac{1}{2}\langle
T_{2}^{2}\rangle _{\mathrm{b}}\approx -\frac{(a-r)^{-3}}{60\pi ^{2}a}%
,\;\langle T_{1}^{1}\rangle _{\mathrm{b}}\approx \frac{(a-r)^{-2}}{60\pi
^{2}a^{2}}.  \label{TiknearCyl}
\end{equation}%
These leading terms do not depend on the planar angle deficit in the cosmic
string geometry. Near the cosmic string the main contribution comes from the
mode $m=0$ and we have%
\begin{equation}
\langle T_{0}^{0}\rangle _{\mathrm{b}}\approx -\langle T_{1}^{1}\rangle _{%
\mathrm{b}}\approx -\langle T_{2}^{2}\rangle _{\mathrm{b}}\approx \frac{q}{%
8\pi ^{2}a^{4}}\int_{0}^{\infty }dx\,x^{3}\left[ \frac{K_{0}(x)}{I_{0}(x)}-%
\frac{K_{1}(x)}{I_{1}(x)}\right] =-0.0168\frac{q}{a^{4}}.  \label{T00nearstr}
\end{equation}%
Therefore, differently from the VEV for the field square, the boundary
induced part in the vacuum energy-momentum tensor is finite on the string
for all values of $q$.

\begin{figure}[tbph]
\begin{center}
\epsfig{figure=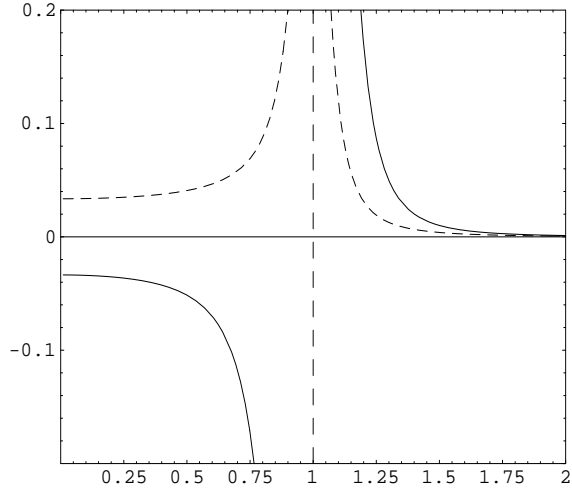,width=7.5cm,height=6.5cm}
\end{center}
\caption{Boundary-induced parts in the VEVs of the energy density (full
curves), $a^{4}\langle T_{0}^{0}\rangle _{\mathrm{b}}$, and the radial
stress (dashed curves), $a^{4}\langle T_{1}^{1}\rangle _{\mathrm{b}}$, as
functions of $r/a$ for the cosmic string with $q=2$.}
\label{fig1}
\end{figure}

\begin{figure}[tbph]
\begin{center}
\epsfig{figure=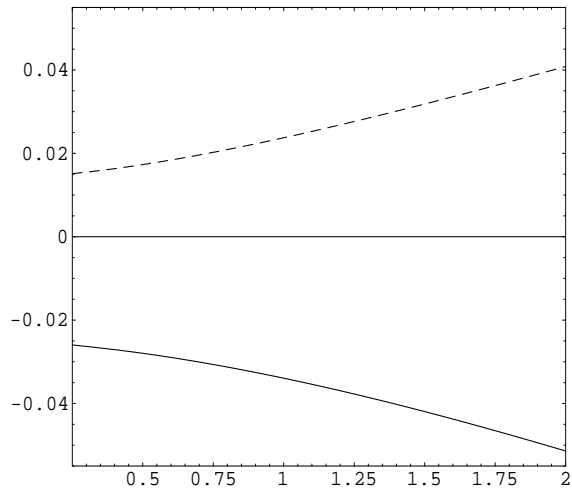,width=7.5cm,height=6.5cm}
\end{center}
\caption{Boundary-induced parts in the VEVs of the energy density (full
curve), $a^{4}\langle T_{0}^{0}\rangle _{\mathrm{b}}$, and the radial stress
(dashed curve), $a^{4}\langle T_{1}^{1}\rangle _{\mathrm{b}}$, evaluated at $%
r/a=0.5$ as functions of the parameter $q$. }
\label{fig2}
\end{figure}

The behavior of the boundary induced part in the VEV of the energy-momentum
tensor in the asymptotic regions of the parameter $q$ is investigated in a
way analogous to that for the field square. For $q\ll 1$, we replace the
summation over $m$ by the integration in accordance with (\ref{replSumInt})
and the VEV tends to a finite limiting value which does not depend on $q$.
Note that as the spatial volume element is proportional to $1/q$, in this
limit the global quantities such as the integrated vacuum energy behave as $%
1/q$. In the limit $q\gg 1$, the contribution of the modes with $m\geqslant 1
$ is suppressed by the factor $\exp [-2qm\ln (a/r)]$ and the main
contribution comes from the $m=0$ mode with the behavior $\propto q$. Though
in this limit the vacuum densities are large, due to the factor $1/q$ in the
spatial volume, the corresponding global quantities tend to finite value. In
figure \ref{fig1} we have plotted the boundary-induced parts in the VEVs of
the energy density (full curves), $\langle T_{0}^{0}\rangle _{\mathrm{b}},$
and radial stress (dashed curves), $\langle T_{1}^{1}\rangle _{\mathrm{b}}$,
as functions of the scaled radial coordinate, $r/a$, for the geometry of the
cosmic string with $q=2$. In figure \ref{fig2} we present the same
quantities evaluated at $r/a=0.5$ as functions of the parameter $q$.

\section{Vacuum densities in the exterior region}

\label{sec:exter}

In this section we consider the VEVs for the field square and the
energy-momentum tensor in the region outside the cylindrical boundary. The
corresponding eigenfunctions for the vector potential are given by formulae (%
\ref{Aalpha}) with the replacement%
\begin{equation}
J_{q|m|}(\gamma r)\rightarrow g_{q|m|}^{(\lambda )}(\gamma a,\gamma
r)=J_{q|m|}(\gamma r)Y_{q|m|}^{(\lambda )}(\gamma a)-Y_{q|m|}(\gamma
r)J_{q|m|}^{(\lambda )}(\gamma a),  \label{extreplace}
\end{equation}%
where, as before, $\lambda =0,1$ correspond to the waves of the electric and
magnetic types, respectively. Now, the eigenvalues for $\gamma $ are
continuous and in the normalization condition the corresponding part on the
right is presented by the delta function. As the normalization integral
diverges for $\gamma ^{\prime }=\gamma $, the main contribution into the
integral comes from large values of $r$ and we can replace the cylindrical
functions with the argument $\gamma r$ by their asymptotic expressions for
large values of the argument. By this way it can be seen that the
normalization coefficient in the exterior region is determined by the
relation%
\begin{equation}
\beta _{\alpha }^{-2}=\frac{2\pi }{q}\gamma \omega \left[ J_{q|m|}^{(\lambda
)2}(\gamma a)+Y_{q|m|}^{(\lambda )2}(\gamma a)\right] .  \label{betalfext}
\end{equation}%
Substituting the eigenfunctions into the corresponding mode-sum formula, for
the VEV of the field square one finds%
\begin{equation}
\langle 0|F^{2}|0\rangle =\frac{q}{\pi }\sideset{}{'}{\sum}_{m=0}^{\infty
}\int_{-\infty }^{+\infty }dk\int_{0}^{\infty }d\gamma \sum_{\lambda =0,1}%
\frac{\gamma ^{3}}{\sqrt{k^{2}+\gamma ^{2}}}\frac{g_{qm}^{(\eta _{F\lambda
})}\left[ k,g_{qm}^{(\lambda )}(\gamma a,\gamma r)\right] }{J_{qm}^{(\lambda
)2}(\gamma a)+Y_{qm}^{(\lambda )2}(\gamma a)},  \label{F2ext}
\end{equation}%
where the functions $g_{qm}^{(\eta _{F\lambda })}\left[ k,g_{qm}^{(\lambda
)}(\gamma a,\gamma r)\right] $ are defined by relations (\ref{gnulam}) with $%
f(x)=g_{qm}^{(\lambda )}(\gamma a,x)$. To extract from this VEV the
boundary-induced part, we subtract from the right-hand side the
corresponding expression for the bulk without boundaries, given by formula (%
\ref{F2s}). The difference can be further evaluated by using the identity%
\begin{equation}
\frac{g_{qm}^{(\eta _{F\lambda })}\left[ k,g_{qm}^{(\lambda )}(\gamma
a,\gamma r)\right] }{J_{qm}^{(\lambda )2}(\gamma a)+Y_{qm}^{(\lambda
)2}(\gamma a)}-g_{qm}^{(\eta _{F\lambda })}\left[ J_{qm}(\gamma r)\right] =-%
\frac{1}{2}\sum_{s=1}^{2}\frac{J_{qm}^{(\lambda )}(\gamma a)}{%
H_{qm}^{(s)(\lambda )}(\gamma a)}g_{qm}^{(\eta _{F\lambda })}\left[
k,H_{qm}^{(s)}(\gamma r)\right] ,  \label{ident}
\end{equation}%
where $H_{qm}^{(1,2)}(z)$ are the Hankel functions. In order to do the
integral over $\gamma $ with the term on the right of (\ref{ident}) we
rotate the integration contour by the angle $\pi /2$ for the term with $s=1$
and by the angle $-\pi /2$ for the term with $s=2$. After introducing the
modified Bessel functions and integrating over $k$ with the help of formula (%
\ref{Hash1}), we can write the VEVs of the field square in the form (\ref%
{F21}), with the boundary induced part given by
\begin{equation}
\langle F^{2}\rangle _{\mathrm{b}}=\frac{q}{\pi }\sideset{}{'}{\sum}%
_{m=0}^{\infty }\sum_{\lambda =0,1}\int_{0}^{\infty }dx\,x^{3}\frac{%
I_{qm}^{(\lambda )}(xa)}{K_{qm}^{(\lambda )}(xa)}G_{qm}^{(\eta _{F\lambda })}%
\left[ K_{qm}(xr)\right] ,  \label{F2bext}
\end{equation}%
where the functions $G_{qm}^{(\eta _{F\lambda })}\left[ f(x)\right] $ are
defined by formulae (\ref{Gnujtilde}). Comparing this result with (\ref{F2b}%
), we see that the formulae for the interior and exterior regions are
related by the interchange $I_{qm}\rightleftarrows K_{qm}$. The
boundary-induced part diverges on the cylindrical shell with the leading
term being the same as that for the interior region. At large distances from
the cylindrical shell we introduce a new integration variable $y=xr$ and
expand the integrand over $a/r$. The main contribution comes from the lowest
mode $m=0$ and up to the leading order we have
\begin{eqnarray}
\langle E^{2}\rangle _{\mathrm{b}} &\approx &\frac{q}{2\pi r^{4}\ln (r/a)}%
\int_{0}^{\infty }dx\,x^{3}\left[ 2K_{0}^{2}(x)+K_{1}^{2}(x)\right] =\frac{2q%
}{3\pi r^{4}\ln (r/a)},  \label{E2far} \\
\langle B^{2}\rangle _{\mathrm{b}} &\approx &-\frac{q}{2\pi r^{4}\ln (r/a)}%
\int_{0}^{\infty }dx\,x^{3}K_{1}^{2}(x)=-\frac{q}{3\pi r^{4}\ln (r/a)}.
\label{B2far}
\end{eqnarray}%
As we see, at large distances the ratio of the boundary-induced and purely
topological VEVs behaves as $1/\ln (r/a)$.

Now we turn to the VEVs of the energy-momentum tensor in the exterior
region. Substituting the eigenfunctions into the corresponding mode-sum
formula, one finds (no summation over $i$)%
\begin{equation}
\langle 0|T_{i}^{k}|0\rangle =\frac{q\delta _{i}^{k}}{8\pi ^{2}}%
\sideset{}{'}{\sum}_{m=0}^{\infty }\int_{-\infty }^{+\infty
}dk\int_{0}^{\infty }d\gamma \sum_{\lambda =0,1}\frac{\gamma ^{3}}{\sqrt{%
k^{2}+\gamma ^{2}}}\frac{f_{qm}^{(i)}\left[ k,g_{qm}^{(\lambda )}(\gamma
a,\gamma r)\right] }{J_{qm}^{(\lambda )2}(\gamma a)+Y_{qm}^{(\lambda
)2}(\gamma a)}.  \label{Tikext0}
\end{equation}%
Subtracting from this VEV the corresponding expression for the bulk without
boundaries, analogously to the case of the field square, it can be seen that
the VEV is presented in the form (\ref{Tik1}), with the boundary-induced
part given by
\begin{equation}
\langle T_{i}^{k}\rangle _{\mathrm{b}}=\frac{q\delta _{i}^{k}}{4\pi ^{2}}%
\sideset{}{'}{\sum}_{m=0}^{\infty }\sum_{\lambda =0,1}\int_{0}^{\infty
}dxx^{3}\frac{I_{qm}^{(\lambda )}(xa)}{K_{qm}^{(\lambda )}(xa)}F_{qm}^{(i)}%
\left[ K_{qm}(xr)\right] ,  \label{Tikbext}
\end{equation}%
where the functions $F_{qm}^{(i)}\left[ f(y)\right] $ are defined by
formulae (\ref{Fnu0}), (\ref{Fnui}). By taking into account the relations
for the modified Bessel functions already used in the previous section and
the inequality $-K_{\nu }^{\prime }(y)>\sqrt{1+\nu ^{2}/y^{2}}K_{\nu }(y)$,
we can see that in the exterior region the boundary-induced vacuum energy
density, axial and radial stresses are positive and the corresponding
azimuthal stress is negative. Note that, in general, the values of the angle
deficit in the exterior and interior regions might be different. In
particular, we can consider a model with zero angle deficit in the interior
region, for which the metric is regular everywhere.

The leading divergence in the boundary induced part (\ref{Tikbext}) on the
cylindrical surface is given by the same formulae as for the interior
region. For large distances from the shell one has%
\begin{equation}
\langle T_{0}^{0}\rangle _{\mathrm{b}}\approx \langle T_{1}^{1}\rangle _{%
\mathrm{b}}\approx -\frac{1}{3}\langle T_{2}^{2}\rangle _{\mathrm{b}}\approx
\frac{q}{24\pi ^{2}r^{4}\ln (r/a)}.  \label{Tikfar}
\end{equation}%
For small values of the parameter $q$, the main contribution to the
boundary-induced VEVs in the exterior region comes from large values of $m$
and the leading terms in the corresponding asymptotic expansion are obtained
from (\ref{F2bext}) and (\ref{Tikbext}) by making use of the replacement (%
\ref{replSumInt}). These terms do not depend on the parameter $q$. For large
values of $q$, $q\gg 1$, the contribution of the terms with $m>0$ is
suppressed by the factor $\exp [-2qm\ln (r/a)]$ and the main contribution
comes form the $m=0$ term with the behavior $\langle F^{2}\rangle _{\mathrm{b%
}}\propto q$ and $\langle T_{i}^{k}\rangle _{\mathrm{b}}\propto q$. In
figure \ref{fig1} we have presented the boundary-induced parts in the vacuum
energy density (full curves) and radial stress (dashed curves) in the
exterior region as functions of $r/a$ for the cosmic string with $q=2$.

\section{Conclusion}

\label{sec:Conc}

In this paper we have investigated the polarization of the electromagnetic
vacuum by a conducting cylindrical shell coaxial with the cosmic string. In
section \ref{sec:Inter} we have considered the interior region. The
corresponding mode-sums for both field square and the energy-momentum tensor
contain series over the zeros of the Bessel function and its derivative. For
the summation of these series we used a variant of the generalized
Abel-Plana formula. The latter allows us to extract from the VEVs the parts
corresponding to the cosmic string geometry without a cylindrical shell and
to present the part induced by the shell in terms of exponentially
convergent integrals for points away from the boundary. In particular, we
have shown that the boundary-free parts of the squares for the electric and
magnetic fields are the same. The corresponding energy-momentum tensor is
obtained from these quantities and coincides with the result previously
obtained in \cite{Frol87,Dowk87b}. In the interior region the shell-induced
parts, which in fact also codify the presence of the cosmic string, are
given by formulae (\ref{F2b0}) for the field squares and by (\ref{Tikb}) for
the energy-momentum tensor. The corresponding energy density and the axial
stress are negative for all values of the angle deficit, whereas the radial
and azimuthal stresses are positive. The expressions for the
boundary-induced parts in the VEVs of the field square and the
energy-momentum tensor outside the shell are given by formulae (\ref{F2bext}%
), (\ref{Tikbext}). In this region the vacuum energy density, axial and
radial stresses are positive and the azimuthal stress is negative. At large
distances from the shell, the ratio of the boundary-induced and purely
topological densities decays logarithmically. The dependence of the vacuum
densities in various limiting regions for the parameters, tells us that for
small values of the parameter $q$ the boundary induced VEVs tend to a finite
limiting value. On the other hand, for large values of this parameter, the
contribution into the boundary induced VEVs coming from the modes with $%
m\neq 0$ is exponentially suppressed, whereas the contribution of the lowest
mode $m=0$ is proportional to $q$. Though in this limit the vacuum densities
are large, due to the factor $1/q$ in the spatial volume element, the
corresponding global quantities tend to finite limiting values.

We have considered the idealized geometry of a cosmic string with zero
thickness. A realistic cosmic string has a structure on a length scale
defined by the phase transition at which it is formed. As it has been shown
in Refs. \cite{Alle90,Alle92,Alle96}, the internal structure of the string
may have non-negligible effects even at large distances. Here we note that
when the cylindrical boundary is present, the VEVs of the physical
quantities in the exterior region are uniquely defined by the boundary
conditions and the bulk geometry. This means that if we consider a
non-trivial core model with finite thickness $b<a$ and with the line element
(\ref{ds21}) in the region $r>b$, the results in the region outside the
cylindrical shell will not be changed. As regards to the interior region,
the formulae given above are the first stage of the evaluation of the VEVs
and other effects could be present in a realistic cosmic string. Note that
from the point of view of the physics in the exterior region the conducting
cylindrical surface can be considered as a simple model of superconducting
string core. Superconducting strings are predicted in a wide class of field
theories and they are sources of a number of interesting astrophysical
effects such as generation of synchrotron radiation, cosmic rays, and
relativistic jets.

\section*{Acknowledgments}

AAS was supported by PVE/CAPES Program and in part by the Armenian Ministry
of Education and Science Grant No. 0124. ERBM and VBB thank Conselho
Nacional de Desenvolvimento Cient\'{\i}fico e Tecnol\'{o}gico (CNPq) and
FAPESQ-PB/CNPq (PRONEX) for partial financial support.

\end{document}